\documentclass[pra,twocolumn,showpacs,groupedaddress,superscriptaddress,aps,10pt]{revtex4-1}
\usepackage{bm,graphicx,amsmath}
\usepackage{placeins}
\usepackage{amssymb}
\usepackage{amsmath}
\usepackage{epsfig}
\usepackage{amssymb}
\usepackage{color}
\usepackage[colorlinks,linkcolor=blue,citecolor=blue]{hyperref}
\usepackage{subfigure}

\begin{document}

\title{Polarization tailored novel vector beams based on conical refraction}
\date{\today}

\author{A. Turpin}\email{Corresponding author: alejandro.turpin@uab.cat}
\affiliation{Departament de F\'isica, Universitat Aut\`onoma de Barcelona, Bellaterra, E-080193, Spain,}
\author{Yu. V. Loiko}
\affiliation{Aston Institute of Photonic Technologies, School of Engineering \& Applied Science Aston University, Birmingham, B4 7ET, UK}
\author{A. Peinado}
\affiliation{Departament de F\'isica, Universitat Aut\`onoma de Barcelona, Bellaterra, E-080193, Spain,}
\author{A. Lizana}
\affiliation{Departament de F\'isica, Universitat Aut\`onoma de Barcelona, Bellaterra, E-080193, Spain,}
\author{J. Campos}
\affiliation{Departament de F\'isica, Universitat Aut\`onoma de Barcelona, Bellaterra, E-080193, Spain,}
\author{T. K. Kalkandjiev}
\affiliation{Departament de F\'isica, Universitat Aut\`onoma de Barcelona, Bellaterra, E-080193, Spain}
\affiliation{Conerefringent Optics SL, Avda Cubelles 28, Vilanova i la Geltr\'u, E-08800, Spain}
\author{J. Mompart}
\affiliation{Departament de F\'isica, Universitat Aut\`onoma de Barcelona, Bellaterra, E-080193, Spain}

\begin{abstract}
Coherent vector beams with involved states of polarization (SOP) are widespread in the literature, having applications in laser processing, super-resolution imaging and particle trapping. We report novel vector beams obtained by transforming a Gaussian beam passing through a biaxial crystal, by means of the conical refraction phenomenon. We analyze both experimentally and theoretically the SOP of the different vector beams generated and demonstrate that the SOP of the input beam can be used to control both the shape and the SOP of the transformed beam. We also identify polarization singularities of such beams for the first time and demonstrate their control by the SOP of an input beam. 
\textbf{OCIS}: (260.5430) Polarization; (260.1440) Birefringence; (260.1180) Crystal optics.
\end{abstract}

\date{\today }
%
%
%
\maketitle
\section{Introduction}
\label{intro} 
The state of polarization (SOP) is one of the fundamental signatures of light fields associated with their vectorial nature. In general, at each point in space the dynamics of the electric field vector of a wave can be described by an ellipse. This ellipse is known as polarization ellipse and it is characterized by the orientation of its major axis through the azimuth angle $\Phi \in [0,\pi]$ and by the ellipticity parameter $\beta \in [-\pi/4,\pi/4]$ so that $\tan{\beta}$ is the ratio of the axes of the polarization ellipse. If $\beta=0$ the light field is linearly polarized, while if $\beta=\pm \pi/4$ the SOP will be circular (left handed for `-' and right handed for `+', if we consider an observer looking in the direction from which light is coming). Usually coherent light beams are homogeneously polarized, i.e. the SOP is identical for all points at any transverse plane along the beam propagation. However, there exist light beams possessing non-homogenous polarization, known as vector beams, such as the well known radial or azimuthal polarizations \cite{zhan2009} or even beams with more involved polarization distributions \cite{carnicer2013,chinosCR,krolikowski2010}.
The non-homogeneous polarization distribution of vector beams can lead to singular points where the SOP is exactly circular (C points), lines along which the SOP is linear (L lines) or disclinations where the instantaneous electric field is null \cite{nye1974,dennis2009,dennis2010NatPhysics,Kleckner2013NatPhysics,freund2012,ref24taco}. 
Vector beams have been applied to laser material processing, optical imaging, atomic spectroscopy, and optical trapping (see \cite{zhan2009} and the references therein), among many others.

\begin{figure}[]
\centering
\includegraphics[width= 1 \columnwidth]{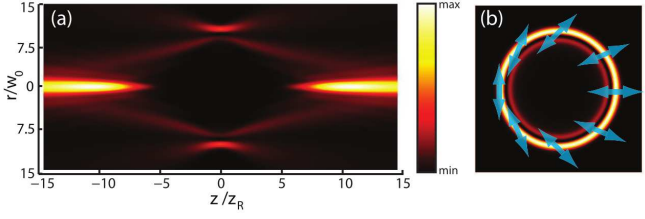}
\caption{CR intensity distribution for a circularly polarized input beam and under conditions of $\rho_0 \equiv R_0 / w_0 \gg 1$. (a) Intensity along the propagation direction z, which possesses cylindrical symmetry. (b) Transverse intensity pattern at $z=0$ (focal plane) showing the two bright rings split by the dark Poggendorff one. Blue double arrows indicate the linear plane of polarization. $w_0$ is the beam waist and $z_R$ the Rayleigh length.}
\label{fig1}
\end{figure}

In recent years, there has been a renewed interest in the conical refraction (CR) phenomenon ocurring in biaxial crystals, which can also provide different vector beams \cite{1978_Belskii_OS_44_436,1999_Belsky_OC_167_1,2004_Berry_JOA_6_289,2007_Berry_PO_55_13,2008_Kalkandjiev_SPIE_6994,ebs2013,multiple-rings2013,2013_Sokolovskii_OE_21_11125,loiko2014,SGCR,hole,phelan2011,peet2010b}. In CR, when a focused Gaussian input beam with waist radius $w_0$ propagates along one of the optic axes of a biaxial crystal it appears transformed at the focal plane as a pair of concentric and well resolved bright rings split by a dark (Poggendorff) ring of geometrical radius $R_0$, as shown in Fig.~\ref{fig1}(b). This intensity distribution is found as long as the condition $\rho_0 \equiv R_0/w_0 \gg 1$ is satisfied \cite{rings_CR, 2004_Berry_JOA_6_289,peinado2013}, while other light structures have been found for $\rho_0 \lesssim 1$ \cite{peet2010b,hole,SGCR}. $R_0$ is the product of the crystal length, $l$, and the CR semi-angle $\alpha$, i.e. $R_0=l \alpha$ \cite{2008_Kalkandjiev_SPIE_6994}. The CR semi-angle $\alpha$ depends on the principal refractive indices of the crystal as ${\alpha = \sqrt{(n_2^2-n_1^2)(n_3^2-n_2^2)/n_2^2}}$, where we have assumed $n_{1}<n_{2}<n_{3}$. 
The non-uniform polarization distribution of a CR beam at the focal plane under Poggendorff splitting conditions is depicted by blue double arrows in Fig.~\ref{fig1}(b). At any point of the CR rings, the electric field is linearly polarized with the azimuth varying so that every pair of diametrically opposite points have orthogonal polarizations. This polarization distribution depends only on the orientation of the plane of optic axes of the biaxial crystal. As a consequence, the intensity distribution along the ring depends on the input SOP, corresponding the one depicted in Fig.~\ref{fig1} to an incident circularly polarized beam. 
For $\rho_0 \gg 1$, the SOP of the CR beam at the focal plane is already well known \cite{2007_Berry_PO_55_13,ebs2013,2013_Sokolovskii_OE_21_11125,peinado2013}. 

Here, for the first time to our knowledge, we investigate the CR SOP out of the focal plane, including values of $\rho_0 \lesssim 1$, when Poggendorff fine splitting vanishes. By means of the Stokes vector formalism, we characterize the resulting novel vector beams of CR and show that the SOP of the input beam can be used to control both the shape and the SOP of the transformed CR beam.
%
%

The article is organized as follows. Section \ref{background} is devoted to briefly review the basics of the Stokes vector and CR formalisms needed for the subsequent sections. In Sections \ref{rings} and \ref{small_rho0},  we report for $\rho_0 \gg 1$ and $\rho_0 \lesssim 1$, respectively, the generation of novel CR vector beams  both theoretically and experimentally. We characterize these vector beams by measuring their Stokes parameters and propose methods to manipulate them as, for instance, making use of the SOP of the input beam.
Finally, we summarize the main results of this work and discuss potential applications of these CR vector beams in different fields of optics. 

\section{Theoretical background}
\label{background}

\subsection{Stokes vector formalism}

The standard tool to analyze the SOP of a light beam is the Stokes vector: $S = (S_0,S_1,S_2,S_3)$. For an electric field $\mathbf{E} = (E_x, E_y)$ with intensity $I$ the Stokes parameters read \cite{ref24taco}:
\begin{eqnarray}
S_0 &=& I = \left| E_x \right|^2 + \left| E_y \right| ^2,\\
\label{S0}
S_1 &=& I_{0^{\circ}} - I_{90^{\circ}} = \left| E_x \right|^2 - \left| E_y \right|^2,\\
\label{S1}
S_2 &=& I_{45^{\circ}} - I_{135^{\circ}} = 2 \rm{Re} \left[ E_x^* E_y \right],\\
\label{S2}
S_3 &=& I_{R} - I_{L} = 2 \rm{Im} \left[ E_x^* E_y \right],
\label{S3}
\end{eqnarray}
where $I_{\Phi}$ ($\Phi=0^{\circ},45^{\circ},90^{\circ},135^{\circ}$) indicates the intensity of linearly polarized light with azimuth $\Phi$, and $I_R$ and $I_L$ indicate the intensity of right- and left-handed circularly polarized light, respectively. In what follows, we will use equations normalized to $E^2$, i.e. we will consider $I = E^2 = 1$. These definitions show that $S_0$ account for the intensity of the light beam, $S_1$ measures the amount of light which is linearly polarized (LP) in the vertical/horizontal basis, $S_2$ does the same but with the diagonal basis and $S_3$ relates the SOP in the right- and left- circularly polarized (CP) basis. The following equations show how the Stokes parameters relate to the azimuth and ellipticity of the polarization ellipse \cite{ref24taco}:
\begin{eqnarray}
\Phi = \frac{1}{2} \arctan{\left( \frac{S_{2}}{S_{1}} \right)},
\label{phi_stokes} \\
\beta = \frac{1}{2} \arctan{\left( \frac{S_{3}}{\sqrt{S_{1}^2+S_{2}^2}} \right)}.
\label{beta_stokes}
\end{eqnarray}

\subsection{Conical refraction theory}

The theoretical model predicting the transformation of an input beam once it passes along one of the optic axis of a biaxial crystal is based on the Belsky-Khapalyuk-Berry (BKB) integrals \cite{1978_Belskii_OS_44_436,2007_Berry_PO_55_13}. For a uniformly polarized input beam with electric field \textbf{E} and cylindrically symmetric 2D Fourier transform ${a \left( \kappa \right)=2\pi\int_{0}^{\infty} r E_{\rm{in}} \left( r \right) J_{0} \left( \kappa r \right) d r}$, the electric field amplitude behind the crystal can be written as follows:
\begin{eqnarray}
&&
\mathbf{E} \left( \rho,Z\right)  =
\left(
\begin{array}{cc}
B_{0} + C & S \\ 
S & B_{0} - C
\end{array}
\right)
\mathbf{e}_{\rm{0}}
,~ \label{Eqs_output_beam_uniform}
\end{eqnarray}
where $C=B_{1} \cos \left( \varphi + \varphi_{C} \right)$ and $S=B_{1} \sin \left( \varphi + \varphi_{C} \right)$. $\mathbf{E}=\left( E_{x}, E_{y} \right)$ denotes the transverse Cartesian components of the output field, $\mathbf{e_0} = \left( e_{\rm{x}}, e_{\rm{y}} \right)$ is the unit vector of the input electric field, and $\varphi_{C}$ is the polar angle that defines the orientation of the plane  of the crystal optic axes. $Z=z/z_{R}$ and $\boldmath{\rho} = \left( \cos \varphi,\sin \varphi \right) r /w_{0}$ define cylindrical coordinates whose origin is associated to the geometrical center of the light pattern at the focal plane ($Z=0$). $z_R = \pi w_0^2 / \lambda$ denotes the Rayleigh length.
Without loosing generality and for simplicity we will consider $\varphi_C = 0$. For any other case, it would be enough to replace $\varphi$ by $\varphi+\varphi_C$ in the following expressions.
$B_{0}$ and $B_{1}$ are the main integrals of the Belsky--Khapalyuk--Berry (BKB) paraxial solution for CR \cite{rings_CR}: 
\begin{eqnarray}
B_{0}(\rho,Z)=\frac{1}{2\pi}\int^{\infty }_{0} \eta a \left( \eta \right) e^{-i \frac{Z}{4} \eta ^{2} } \cos \left( \eta \rho_{0}\right) J_{0}\left( \eta \rho \right) d\eta, 
\label{Bc}
\\B_{1}(\rho,Z)=\frac{1}{2\pi}{\int}^{\infty }_{0} \eta a \left( \eta \right) e^{-i \frac{Z}{4} \eta ^{2} } \sin \left( \eta \rho_{0}\right) J_{1}\left( \eta \rho \right) d \eta,
\label{Bs}
\end{eqnarray}
where $\eta=\kappa w_0$ and $J_{q}$ is the $q^{th}$-order Bessel function of the first type. Below we consider a fundamental Gaussian input beam with normalized transverse profile of the electric field amplitude ${E(\rho)=\sqrt{2P/\pi w(Z)^{2}} \exp (-\rho^2)}$ and 2D Fourier transform ${ a \left( \eta \right)=\sqrt{2P\pi /w(Z)^2}\exp -\eta^{2}/ 4}$, being $w(Z) = w_0^2 \sqrt{1+Z^2}$. 
Note the central role $\rho_0 \equiv R_0/w_0$  in Eqs.~(\ref{Bc}) and (\ref{Bs}) so it will determine the resulting CR intensity distribution. Accordingly, we will use $\rho_0$ as the control parameter for our investigations.
If $\rho_0 \gg 1$ is fulfilled then the transverse intensity pattern at the focal plane of the system is formed by the two characteristic bright rings split by the dark Poggendorff one, as shown in Fig.\ref{fig1}(b). As the imaging plane is moved away from the focal plane, more involved structures including secondary rings are found. At positions given by
\begin{equation}
Z_{\rm{Raman}} = \pm \sqrt{\frac{4}{3}} \rho_0 ,
\label{eqraman}
\end{equation}
a bright spot known as the Raman spot \cite{vault} appears on the beam axis, see Fig.~\ref{fig1}(a).

\section{State of polarization for $\rho_0 \gg 1$}
\label{rings}


The first information that can be extracted from Eqs.~(\ref{Eqs_output_beam_uniform})-(\ref{Bs}) with respect to the SOP of the CR beam is that at $\rho = 0$ there is only contribution of $B_{0}$, since $B_{1} \propto J_{1}(\eta \rho = 0) = 0$. Additionally, from Eqs.~(\ref{Bc}) and (\ref{Bs}) note that the SOP of the 
$B_{0}$ component is $\textbf{e}_0$. As a consequence, the center of the CR beam will possess always the same SOP as the input beam. This fact, that was already pointed out in \cite{phelan2009,peet2010a,peet2013}, will be discussed with more detail below. To obtain the Stokes parameters of the CR beam, Eqs.~(\ref{S0})-(\ref{S3}) must be combined with Eqs.~(\ref{Eqs_output_beam_uniform})-(\ref{Bs}). 
For a CP input beam, the electric field and intensity beyond the crystal become:
\begin{eqnarray}
E_{x} &=& B_{0} + B_{1} e^{\pm i \varphi},\\
\label{Ex_CP}
E_{y} &=& \pm i B_{0} \mp i B_{1} e^{\pm i \varphi},\\
\label{Ey_CP}
I_{\rm{CP}} &=& 2 (\left| B_{0} \right|^{2} + \left| B_{1} \right|^{2})
,~
\label{Eqs_output_beam_intensity_CP}
\end{eqnarray}
where upper/lower sign stays for LHCP/RHCP beam. 
For a LP input beam, the corresponding electric field and intensity beyond the crystal read as follows:
\begin{eqnarray}
E_{x} &=& B_{0} \cos \Phi + B_{1} \cos \left( \varphi - \Phi \right),\\
\label{Ex_LP}
E_{y} &=& B_{0} \sin \Phi + B_{1} \sin \left( \varphi - \Phi \right),\\
\label{Ey_LP}
I_{\rm{LP}} &=& I_{\rm{CP}} + 2 {\rm Re} \left[ B_{0} B_{1}^{*} \right]
\cos \left( 2 \Phi - \varphi \right),~ 
\label{Eqs_output_beam_intensity_LP}
\end{eqnarray}
where $\Phi$ is the polarization azimuth of the LP input beam with $\mathbf{{e}_{0}}= \left( \cos \Phi , \sin \Phi \right)$. 

For well resolved concentric rings with Poggendorff splitting, i.e. for $\rho_0 \gg 1$, Eqs.~(\ref{Bc}), (\ref{Bs}), (\ref{Eqs_output_beam_intensity_CP}) and (\ref{Eqs_output_beam_intensity_LP}) show that a radially symmetric intensity pattern of CR is obtained only for a CP input beam. Instead, for a LP input beam, a crescent annular intensity pattern appears such that the zero intensity point is obtained for the ring position that possesses orthogonal polarization to the input beam. In both cases, the polarization distribution of the CR pattern is the same. Every point of the rings is linearly polarized and the azimuth rotates continuously along the ring so that every two diametrically opposite points have orthogonal polarizations. 

\begin{figure}[htbp]
\centering
\includegraphics[width= 1.0 \columnwidth]{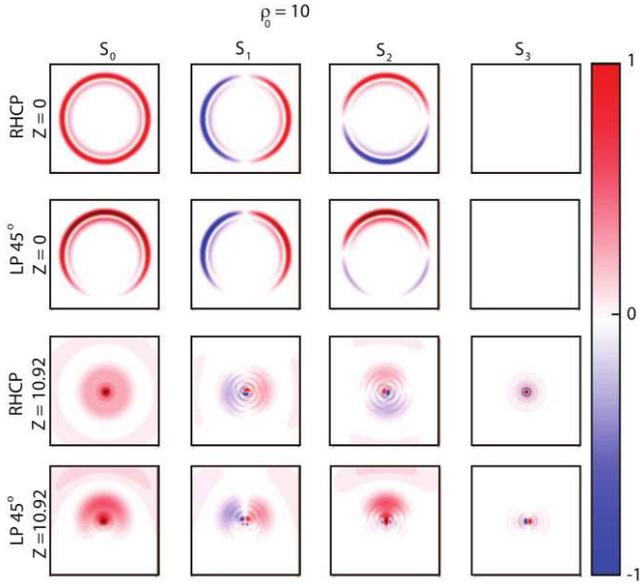}
\caption{Theory: transverse pattern for $\rho_0=10$ of the Stokes parameters $S_{0},S_{1},S_{2},S_{3}$ obtained from numerical simulations for the CR beam transverse profile with a RHCP and a LP ($\Phi = 45^{\circ}$) Gaussian input beam.
First and second rows correspond to the focal plane ($Z = 0$) while third and fourth rows to the Raman spot plane ($Z = 10.92$). The plane of optic axes of the crystal lies horizontally ($\varphi_{c}=0$).}
\label{fig2}
\end{figure}

Fig.~\ref{fig2} presents the numerically obtained Stokes parameters at transverse sections of the CR beam ($\rho_0 = 10$) at $Z=0$ (first and second rows) and $Z=10.92$ (third and fourth rows) obtained from a RHCP (first and third rows) and a LP ($\Phi = 45^{\circ}$) (second and fourth rows) Gaussian input beam. At the focal plane, see first two rows in Fig.~\ref{fig2}, the SOP described by the Stokes parameters is the expected: symmetric pattern for the RHCP case and with a node at $\varphi = 270^{\circ}$ ($\Phi = 135^{\circ}$, since we have used $\varphi_C = 0^{\circ}$). 
Last column in Fig.~\ref{fig2} demonstrates that at the focal plane the SOP of the CR beams, either RHCP or LP ($\Phi = 45^{\circ}$) (or any other) is linear, i.e. $S_{3} = 0$. 
In contrast the Raman spot, $Z = 10.92$, all Stokes parameters are substantially different from zero, as shown in the last two rows in Fig.~\ref{fig2}.

Stokes parameters in Fig.~\ref{fig2} clearly identify polarization singularities of CR beams. 
For RHCP input light, it is a C-point at the center of the CR beam, i.e. it is of circular polarization at any point of the beam center along propagation. The center of the $S_{3}$ transverse pattern is a point with maximum intensity, while the other two Stokes parameters $S_{1}$ and $S_{2}$ have zero values. 
For LP$_{45^{\circ}}$ input light, one can identify L-line singularity. In Fig.~\ref{fig2} this line can be identified as a vertical line of zero value of the Stokes parameters $S_{3}$ and $S_{1}$ and nonzero value of $S_{2}$. It belongs to the plane defined by the points of CR ring with linear SOP of LP$_{\Phi = 45^{\circ}}$ and LP$_{\Phi = 135^{\circ}}$.

\href{media1.avi}{Supplement 1} and \href{media2.avi}{Supplement 2} show, respectively, the spatial evolution of the Stokes parameters along the propagation direction for a RHCP and LP ($\Phi = 45^{\circ}$) Gaussian input beam. As we move away from the focal plane, the bright rings become wider and the intensity at the Poggendorff dark ring is no longer zero. The intensity of the outer ring decreases while the intensity of the inner ring increases. The outer ring expands and the inner ring becomes smaller in radius. As a result, at $Z \approx 6$ a spot in the center of the pattern appears. Finally, the inner ring shrinks into a bright spot at $Z = 10.92$ for $\rho_0 = 10$, corresponding to the Raman spot. 
At this point, there can be found an additional type of polarization singularity independently of the polarization state of an input beam. They are clearly distinguishable by expecting Stokes parameter $S_{3}$ that defines degree of circular polarization of the field. Inspection of $S_{3}$ far from the focal plane, for instance, at the Raman spot as demonstrated in the last two rows in Fig.~\ref{fig2}, reveals alternating annular regions of RHCP and LHCP states. These annular regions of circular polarization are separated by circles of null intensity. At these circles the field is linearly polarized. Therefore, these polarization singularities can be called as L-circles. This behavior has been also observed in the focusing of radially polarized beams and explored in detail in Ref. \cite{visser2006}. 
For a LP ($\Phi = 45^{\circ}$) input beam, the central spot is broken by a line of null intensity (L-line) that connects the two points with LP and azimuth $\Phi = 45^{\circ}$ and $\Phi = 135^{\circ}$, as commented before. 

One feature that supplements demonstrate is that polarization singularities, i.e. C-point for CP input beam, L-lines for LP input beam and L-circles, are invariants of CR beam propagation behind the crystal.
Another feature demonstrated by the supplement movies is the inversion of the center of the $S_{1}$ and $S_{2}$ parameters after the focal plane, which can be understood as a manifestation of the Gouy phase \cite{boyd1980,visser2013} of the CR beam. This effect will be more appreciable in the CR vector beams discussed in the following section. 

\begin{figure}[htbp]
\centering
\includegraphics[width= 1.0 \columnwidth]{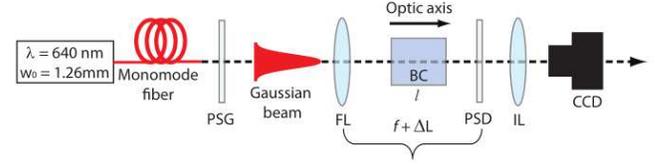}
\caption{Experimental set-up. A diode laser coupled to a monomode fiber generates a Gaussian beam at $640\,\rm{nm}$ with a beam waist radius $w_0 = 1.26\,\rm{mm}$. Then the beam is focused by means of a focusing lens (FL) along one of the optic axes of a $\rm{KGd(WO_4)_2}$ biaxial crystal (BC). Experiments from Fig.~\ref{rho0_10_exp} were carried out using a FL with $100\,\rm{mm}$ focal length and a biaxial crystal $10.5\,\rm{mm}$ long, while FLs with focal lengths of $150\,\rm{mm}$, $200\,\rm{mm}$ and $400\,\rm{mm}$ and a biaxial crystal $2.3\,\rm{mm}$ long were used for the experiments from Fig.~\ref{stokes_exp}. Linear and circular polarizers are used as polarization state generators (PSG) and polarization state detectors (PSD) to generate and measure the SOP of the input and output beam, respectively. The transverse patterns are recorded by means of an imaging lens (IL) that projects the image into a CCD camera.}
\label{fig5}
\end{figure}

To test the validity of the obtained theoretical results, we have performed the corresponding experimental measurements. Fig.~\ref{fig5} shows the experimental set-up. The Gaussian input beam is obtained from a diode laser at $640\,\rm{nm}$ coupled to a monomode fiber with a collimator, yielding a beam waist of $w_0 = 1.26\,\rm{mm}$. 
To generate the different input polarization states (LP with $\Phi = 45^{\circ}$ and RHCP) we use a $\lambda/2$ and a $\lambda/4$ waveplates. We use lenses with different focal lengths to modify the waist radius $w_0$ of the input beam and to reach different values of $\rho_0$. The beam passes along one of the optic axes of a biaxial crystal, placed always before the expected focal plane of the beam. An additional imaging lens is used to take different planes along the beam propagation direction and transfer them onto the CCD camera. Linear and circular polarizers are used to measure the Stokes parameters of the beam after being transformed by the CR phenomenon. We use a commercially available (CROptics) $\rm{KGd(WO_4)_2}$ biaxial crystals with $\alpha = 16.9$~mrad and length $l = 10.5\,\rm{mm}$ yielding CR ring radius of $R_0 = 180\,$$\mu$m. For more details about the material and the alignment procedure, see e.g. Ref.~\cite{SGCR}. 

\begin{figure}[htbp]
\centering
\includegraphics[width= 1.0 \columnwidth]{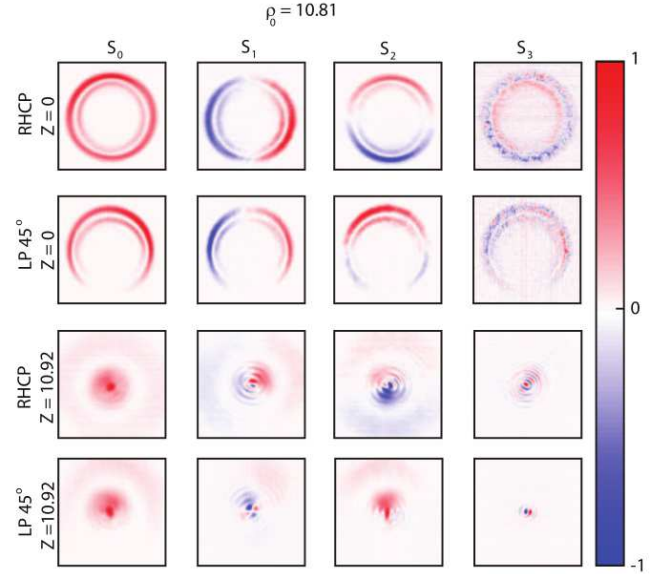}
\caption{Experiment: for $\rho_0=10.81$, transverse patterns of the Stokes parameters $S_{0},S_{1},S_{2},S_{3}$ for the CR beams measured with a RHCP and a LP ($\Phi = 45^{\circ}$) Gaussian input beam. First and second rows correspond to the focal plane ($Z = 0$) while third and fourth rows to the Raman spot plane ($Z = 10.92$).}
\label{rho0_10_exp}
\end{figure}

Fig.~\ref{rho0_10_exp} shows the obtained experimental Stokes parameters for $\rho_0 = 10.81$.
The experimental results are in good agreement with the theoretical results presented in Fig.~\ref{fig2}. Discrepancy has been observed only for the $S_{3}$ parameter for the case of a LP input beam (see last image of second rows). This can be explained in terms of the experimental error introduced by the polarization state detector elements used, that disturb the beam shape and its position, which is central for the quality of the experimental results. 

\section{State of polarization for $\rho_0 \lesssim 1$}
\label{small_rho0}
\begin{figure}[]
\centering
\includegraphics[width= 1 \columnwidth]{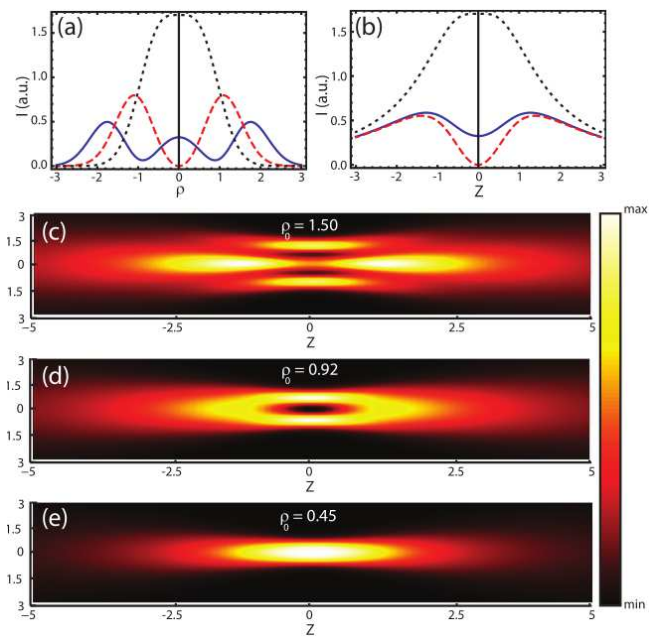}
\caption{Intensity variation (a) along the radial direction $\rho$ at the focal plane $Z=0$ and (b) along the axial direction $Z$ at the beam center ($\rho =0$) for CR vector beams obtained using $\rho_0 = 1.50$ (blue-solid line), $\rho_0 = 0.92$ (red-dashed line) and $\rho_0 = 0.45$ (black-dotted line). The corresponding intensity distribution in the $\left( Z, \rho \right)$ plane are shown in figures (c)-(e)
}
\label{fig3}
\end{figure}

For $\rho_0 \lesssim 1$ CR patterns are significantly different from the double bright concentric rings with clear Poggendorff splitting (occurring for $\rho_0 \gg 1$ as shown in the previous section).
The region $\rho_0 \lesssim 1$ has been explored recently \cite{peet2010b, hole, SGCR} showing that CR can be used to new CR lasers \cite{loiko2014}, to increase the directivity of laser beams \cite{peet2010b}, to generate a super-Gaussian beam \cite{SGCR}, to create a three dimensional dark focus \cite{hole} and even to develop a novel scheme for super-resolution microscopy \cite{sirat2013a,sirat2013b}.
However, in all these works, no deep insight about the SOP of the generated CR beams has been provided. In what follows by considering the Stokes parameters we uncover the evolution of the SOP and polarization singularities of the CR beams and demonstrate how they depend on the SOP of the input beam. Fig.~\ref{fig3} shows the main features and general view of CR beams with $\rho_0 = [1.50,0.92,0.45]$. 
The cross-section of the CR transverse intensity pattern at the focal plane and far away from the focal plane are shown in Figs.~\ref{fig3}(a) and (b), respectively. Figs.~\ref{fig3}(c)-(e) are 2D density plots of the intensity of the CR beams in the $Z$--$\rho$ plane.

%

\begin{figure*}[htbp]
\centering
\includegraphics[width= 2 \columnwidth]{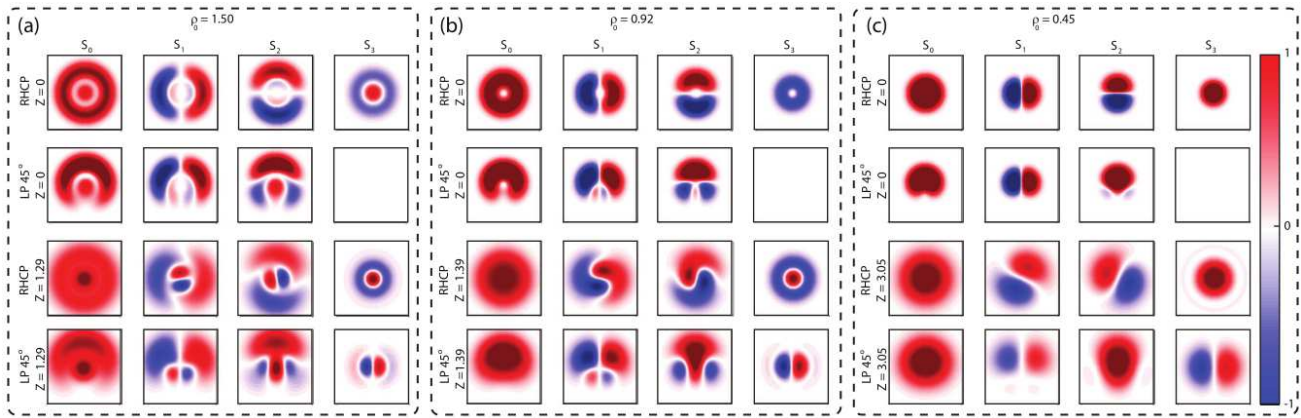}
\caption{Numerically calculated Stokes parameters for $\rho_0 = 1.50$ (a), $\rho_0 = 0.92$ (b), and $\rho_0 = 0.45$ (c). 
The plane of optic axes of the crystal lies horizontally ($\varphi_{c}=0$).}
\label{fig4}
\end{figure*}

Fig.~\ref{fig4} presents density plots of the numerically calculated Stokes parameters for (a) $\rho_0 = 1.50$, (b) $\rho_0 = 0.92$ and (c) $\rho_0 = 0.45$ at $Z=0$ (first and second rows) and out of the focal plane (third and fourth rows). For $\rho_0 = 1.50$ and $\rho_0 = 0.92$ these planes give the axial intensity maximum, while for $\rho_0 = 0.45$ we have considered the plane where the cross-section area of the beam is doubled \cite{SGCR}. First and third rows refer to a RHCP Gaussian input beam while second and fourth rows present the case of a LP ($\Phi=45^{\circ}$) Gaussian input beam. For $\rho_0 \lesssim 1$ the position of the Raman spots are not well determined by Eq.~(\ref{eqraman}) and the position of the axial intensity maxima must be determined for each particular case of $\rho_0$. Its value ($Z$) is indicated at each image. 

While transverse distribution for intensity and polarization of CR beam look different for the $\rho_0 \lesssim 1$ case with respect to $\rho_0 \gg 1$, it should be noted that singularities, i.e. topological structure of the CR beam, remains similar. It means that (i) the CR beam center constitutes a C-point for the case of RHCP input beam (see the first and third rows in Fig.~\ref{fig4}), (ii) there is L-line polarization singularity for the LP$_{\Phi = 45^\circ}$ input beam (see the second and fourth rows in Fig.~\ref{fig4}) and (iii) there are L-circle singularity for input beam with arbitrary SOP (see the third and fourth rows in Fig.~\ref{fig4}).

If we consider the special case of $\rho_0 = 0.92$, the central point at the focal plane is a null-intensity point and therefore all Stokes parameters are also 0 at the beam center. Vanishing intensity at the beam center leads to another feature. Namely, sign of circular polarization state associated with Stokes parameter $S_{3}$ is reversed at the focal plane with respect to the input beam. It means that at the focal plane the CR beam is predominantly of LHCP in the case of RHCP of the input beam we consider in this paper (see first row in Fig.~\ref{fig4} for the case of $\rho_0 = 0.92$). 
Mathematically, it is defined by zero of the integral $B_0$, which leads to dominance of the contribution of integral $B_1$ associated with SOP orthogonal to the input one.
For the case of a LP ($\Phi = 45^{\circ}$) input beam and for all values of $\rho_0$ investigated, at the focal plane $Z=0$ the transverse patterns are crescent-like, with the point of intensity minimum at a point, which is diagonally opposite to the point with maximum intensity. In the case of LP$_{\Phi = 45^{\circ}}$ input beam that we consider, the intensity minimum is observed at the bottom, corresponding to the point of LP with $\Phi = 135^{\circ}$. In contrast to the RCHP case, for all three values of $\rho_0$ the $S_{3}$ parameter is null, which means that the patterns are completely LP. Out of the focal plane $S_{3} \neq 0$ except at the beam center, where there is an L-line connecting the points with LP $\Phi = 45^{\circ}$ and $\Phi = 135^{\circ}$. 

Additional features of focused CR beams are associated with Gouy phase. It can be revealed by considering evolution of transverse pattern for Stokes parameters along CR beam propagation shown in \href{media3.avi}{Supplement 3}--\href{media8.avi}{Supplement 8} for a RHCP and LP ($\Phi = 45^{\circ}$) Gaussian input beam for $\rho_0 = 1.50$, $\rho_0 = 0.92$ and $\rho_0 = 0.45$. 
For RHCP input light (\href{media3.avi}{Supplement 3}, \href{media5.avi}{Supplement 5}, \href{media7.avi}{Supplement 7}) the focal plane ($Z=0$) is a symmetry plane for $S_{3}$. In contrast, $S_{1}$ and $S_{2}$ are rotated roughly $180^{\circ}$ before and after the focal plane, which must be associated to the Gouy phase \cite{boyd1980,visser2013}. For a LP ($\Phi = 45^{\circ}$) input beam (\href{media4.avi}{Supplement 4}, \href{media6.avi}{Supplement 6}, \href{media8.avi}{Supplement 8}) the Stokes parameters $S_{1}$ and $S_{2}$ are symmetric with respect to the focal plane and now $S_{3}$ suffers from a phase shift of $180^{\circ}$, due to the Gouy phase too.  For deeper insight into the Gouy phase effects on the polarization distribution in focused beams, please see Ref.~\cite{visser2013}.

%


%
\begin{figure*}[htbp]
\centering
\includegraphics[width= 2 \columnwidth]{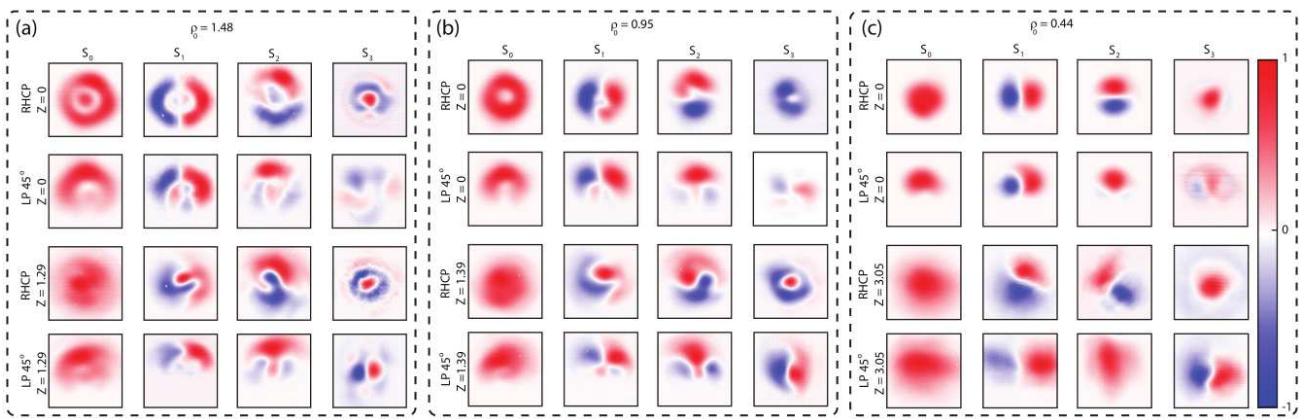}
\caption{Measured Stokes parameters for: (a) $\rho_0 = 1.48$, (b) $\rho_0 = 0.95$, and (c) $\rho_0 = 0.44$.}
\label{stokes_exp}
\end{figure*}

We have performed experiments in order to prove theoretical findings reported above. Fig.~\ref{stokes_exp} shows the experimentally measured Stokes parameters for $\rho_0 = 1.48$ (box (a)), $\rho_0 = 0.95$ (box (b)) and $\rho_0 = 0.44$ (box (c)). For these experiments, the same set-up shown in Fig.~\ref{fig5} was used but in this case taking a $2.3\,\rm{mm}$ long $\rm{KGd(WO_4)_2}$ biaxial crystal yielding CR ring radius of $R_0 = 39$~$\mu$m.
In general, the theoretical predictions are in good agreement with the experimental results. Again, the $S_{3}$ parameter for a LP input beam (last image of second and fourth rows) is the measurement that mostly differs with respect to the numerical predictions. In addition to the experimental difficulties commented above, here it must be also taken into account the fact that small changes in the $\rho_0$ can modify quantitatively the CR pattern. 


\section{conclusions}
\label{conclusions}
In summary, we have studied in detail the SOP of conical refraction based vector beams for a wide range of $\rho_0$ and for different SOP of the input beam. We have determined the Stokes parameters of the CR beam at different transverse planes along the beam propagation direction and we have shown that both the shape and the SOP of the transformed beams depend on the SOP of the input beam. We have shown that the polarization distribution formed by orthogonal polarizations at any two radially opposite points of the pattern and usually associated to the CR phenomenon remains relevant under the condition $\rho_0 \gg 1$ and at the focal plane only. 
For $\rho_0 \lesssim 1$ and CP input beams we have found that CR beams demonstrate more involved structure of
non-homogeneously elliptically polarized states not only with different azimuth but also with different ellipticity. In contrast, for LP input beams, the SOP of the CR has been reported to be completely linear at the focal plane and with variable ellipticity and azimuth out of it. A good agreement between the theoretical predictions and the experimental results has been obtained. 

Additionally, we demonstrated for the first time to our knowledge, experimental results on polarization singularities of CR beams and demonstrate how types of polarization singularities can be changed by and controlled with varying SOP of the input light beams. Such polarization singularities as C-points, L-lines and
L-circles have been identified for CR beams.

The reported results can be particularly interesting for experiments with tightly focused beams \cite{zhan2009}, for the generation of novel polarizations in CR \cite{phelan2011}, in optical micromanipulation \cite{vault,phelan2009b,krolikowski:2014:np}, mode conversion between Heremite--Gauss-like beams and Laguerre--Gauss-like beams \cite{peetLG} and in super-resolution imaging \cite{sirat2013a,sirat2013b}.
It is also promising the generation of polarization-tunable potentials to inject, extract and manipulate ultra-cold atoms \cite{rings_CR, loiko2014b}. Additionally, by taking into account that the CR beams posses orbital angular momentum \cite{berry:2005:jo} and that the biaxial crystals used are transparent to a wide frequency range, the presented technique could be an alternative for the production of high-frequency vortex beams for molecular-scale super-resolution microscopy \cite{boyd:2014:prl}. 
Finally, by using quantum sources, CR vector beams can be a tool for testing new Bell-like inequalities with hybrid polarization-momentum states useful in quantum information technologies \cite{zeilinger:2014:pra}.  


\section*{Funding Information}
Spanish Ministry of Science and Innovation (MICINN) (contracts FIS2011-23719 and FIS2012-39158-C02-01). A.T. acknowledges financial support through the grant AP2010-2310.


\end{document}